\documentclass[twocolumn,showpacs,prc]{revtex4}
\usepackage{graphicx}
\usepackage{dcolumn}
\usepackage{bm}
\begin{document}

\title{Neutron Production from the Fracture of Piezoelectric Rocks}
\author{A. Widom and J. Swain}
\affiliation{Physics Department, Northeastern University, Boston MA, USA}
\author{Y.N. Srivastava}
\affiliation{Department of Physics \& INFN, University of Perugia, Perugia, IT}

\begin{abstract}
A theoretical explanation is provided for the experimental evidence that  
fracturing  piezoelectric rocks produces neutrons. The elastic energy micro-crack 
production ultimately yields the macroscopic fracture. The mechanical energy is 
converted by the piezoelectric effect into electric field energy. The electric field 
energy decays via radio frequency (microwave) electric field oscillations. The 
radio frequency  electric fields accelerate the condensed matter electrons which then 
collide with protons producing neutrons and neutrinos.

\end{abstract}


\maketitle

\section{Introduction \label{intro}}

There has been considerable evidence of high energy particle production during the 
fracture of certain kinds of 
crystals\cite{Karassey:1953, Klyuev:1986, Klyuev:1987, Preparata:1991, Nakayama:1992, Lawn:1993}. 
In particular, fracture induced nuclear transmutations and the production of 
neutrons have been clearly 
observed\cite{Derjaguin:1989, Lipson:1989, Derjaguin:1990, Kaushik:1997, Shioe:2009, Cardone:2009, Carpinteri:2009, Fuji:2002}.   
The production of neutrons appears greatly enhanced if the solids being fractured are 
piezoelectric\cite{Landau:1984} materials. Our purpose is to describe theoretically  the manner 
in which the mechanical pressure in a piezoelectric stressed solid about to fracture can organize 
the energy so that neutrons can be produced. 

The nuclear physics involves a standard weak interaction wherein collective radiation 
plus an electron can be captured by a proton to produce a neutron plus a neutrino
\begin{equation}
({\rm radiation\ energy})+e^- + p^+ \to n + \nu_e .
\label{intro1}
\end{equation}
The required collective radiation energy may be produced by the mechanical elastic energy storage 
via the piezoelectric effect. By the {\em definition} of a {\em piezoelectric material}, the conversions of energy of the form
\begin{equation}
({\rm elastic\ energy})\ \  \Longleftrightarrow \ \  ({\rm electric\ energy})
\label{intro2}
\end{equation}
are allowed.

\begin{figure}
\scalebox {0.6}{\includegraphics{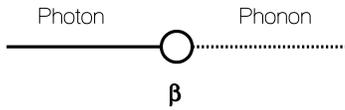}}
\caption{Shown is the Feynman diagram exhibiting the change of a phonon described by the 
tensor strain ${\sf w}$ into a photon described by the vector electric field ${\bf E}$ 
and vice versa. The piezoelectric coupling strength tensor $\beta_{i,jk}$ is exhibited in 
the interaction  
Hamiltonian Eq.(\ref{intro3})} 
\label{fig1}
\end{figure}

In terms of the electric field \begin{math} {\bf E} \end{math} and the crystal strain tensor 
\begin{math} {\sf w} \end{math}, the precise definition of the piezoelectric tensor 
\begin{math} \beta \end{math} is discussed in Sec.\ref{pi}. The final result may be expressed 
as the effective interaction Hamiltonian 
\begin{equation}
{\cal H}_{\rm int}=-\int \beta_{i,jk}E_i w_{jk} d^3{\bf r},
\label{intro3}
\end{equation}
wherein the tensor coefficients \begin{math} \beta_{i,jk} \end{math} describe 
piezoelectricity as shown in FIG. \ref{fig1}. Some implications of the conversion 
from mechanical energy into electromagnetic energy are quite striking. For example, 
a piezoelectric ignition system can be constructed wherein a sharp mechanical 
impulse to a piezoelectric material can induce a sharp voltage spike across the 
sample with the resulting spark igniting a fire in a surrounding gas. More dramatically, 
the rocks crushed in earthquakes contain piezoelectric quartz. The mechanical impulse 
causing micro-cracks in the rocks can thereby produce impulse earthquake lightning flashes. 

In Sec.\ref{fs} we review the stresses and strains which accompany micro-cracks in 
rocks that are being fractured. Elasticity theories of such micro-cracks are well 
known\cite{Fruend:1998, Landau:1970, Griffith:1921}. The central result is as follows.  
If \begin{math} \sigma_{\rm bond} \end{math} denotes the elastic stress required to 
break the chemical bonds on an area of a micro-crack and 
\begin{math} \gamma_{\rm s} \end{math} denotes the surface tension 
of the free face of a crack, then the fracture stress 
\begin{math} \sigma_F  \end{math} required to create a crack of half length 
\begin{math} a \end{math} is given by 
\begin{equation}
\sigma_F=\sqrt{\frac{\sigma_{\rm bond}\gamma_{\rm s}}{a}} 
\ \ \Rightarrow \ \ \sigma_F\ll \sigma_{\rm bond} 
\label{intro4}
\end{equation}
for brittle fracture.

In Sec.\ref{np}, the manner in which the conversion of mechanical to electrical energy 
takes place is explored. It is shown that copious electromagnetic energy is emitted in 
the radio frequency microwave regime. The radiation accelerates the electrons allowing 
for nuclear transmutations in forms following from Eq.(\ref{intro1}). In the concluding 
Sec.\ref{conc}, the number of neutrons produced by rock fractures will be estimated.    

\section{Piezoelectric Interactions \label{pi}}

The energy per unit volume \begin{math} U \end{math} of a piezoelectric material obeys 
\begin{equation}
dU=TdS+{\sf \sigma}:d{\sf w}-{\bf P}\cdot d{\bf E}, 
\label{pi1}
\end{equation}
wherein \begin{math} T \end{math} is the temperature, \begin{math} S \end{math} is the 
entropy, \begin{math} {\sf \sigma} \end{math} is the stress tensor, 
\begin{math} {\sf w} \end{math} is the strain tensor, \begin{math} {\bf P} \end{math} 
is the electric dipole moment per unit volume and \begin{math} {\bf E} \end{math} is 
the electric field. The adiabatic piezoelectric tensor may be defined as 
\begin{equation}
\beta_{i,jk}=\left(\frac{\partial P_i}{\partial w_{jk}}\right)_{S,{\bf E}}
=-\left(\frac{\partial \sigma_{jk}}{\partial E_i}\right)_{S,{\sf w}}.
\label{pi2}
\end{equation}
To quadratic order, the mechanical electric field interaction energy 
\begin{math} U_{int} \end{math} follows from Eq.(\ref{pi2}); It is 
\begin{eqnarray}
\beta_{i,jk}=-
\frac{\partial^2 U(S,{\sf w},{\bf E})}{\partial E_i \partial w_{jk}}=-
\frac{\partial^2 U(S,{\sf w},{\bf E})}{\partial w_{jk} \partial E_i}\ ,
\nonumber \\ 
U_{int}=-\beta_{i,jk}E_i w_{jk} +\ldots \  ,
\label{pi3}
\end{eqnarray}
leading to the quantum operator in the effective Hamiltonian and the Feynman 
diagram of Eq.(\ref{intro3}).  

\begin{figure}
\scalebox {0.6}{\includegraphics{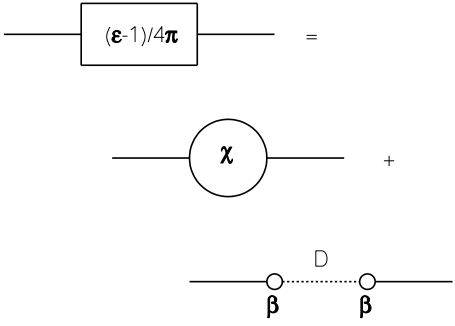}}
\caption{The Feynman diagrams contributing to the polarization part of the photon 
propagator in a piezoelectric material are shown above. The resulting dielectric 
response in Eqs.(\ref{pi7}) and (\ref{pi8}), i.e. $\varepsilon_{ij}(\zeta )$, 
has a contribution due to mechanical phonon modes as exhibited above in 
diagrammatic form.} 
\label{fig2}
\end{figure}

The adiabatic electric susceptibility of the material at constant strain is 
defined 
\begin{equation}
\chi_{ij}=\left(\frac{\partial P_i}{\partial E_j}\right)_{S,{\sf w}},
\label{pi4}
\end{equation} 
while the same susceptibility at constant stress is given by 
\begin{equation}
\tilde{\chi}_{ij}=\left(\frac{\partial P_i}{\partial E_j}\right)_{S,{\sf \sigma }},
\label{pi5}
\end{equation} 
The elastic response tensor, 
\begin{equation}
D_{ijkl}=
\left(\frac{\partial w_{ij}}{\partial \sigma_{kl}}\right)_{S,{\bf E}},
\label{pi6}
\end{equation}
determines the difference between between the two susceptibilities in 
Eq.(\ref{pi4}) and (\ref{pi5}); i.e. the thermodynamic identity is that   
\begin{equation}
\tilde{\chi}_{ij}=\chi_{ij}+\beta_{i,lk}D_{lknm}\beta_{j.nm}.
\label{pi7}
\end{equation}
For a complex frequency \begin{math} \zeta =\omega +i\eta   \end{math} with 
\begin{math} \eta ={\Im }m\ \zeta \ge 0  \end{math}, there are dynamical electric 
susceptibilities \begin{math} \tilde{\chi }_{ij}(\zeta ) \end{math} and 
\begin{math} \chi_{ij}(\zeta ) \end{math}. The dynamical version of Eq.(\ref{pi7}) 
is easily obtained. Phonon modes described by the dynamical phonon propagator 
\begin{math} D_{lknm}(\zeta ) \end{math} affect the dynamical susceptibilities via 
\begin{eqnarray}
{\bf D}={\bf E}+4\pi {\bf P},
\nonumber \\ 
\varepsilon_{ij}(\zeta )=\delta_{ij} +4\pi \tilde{\chi}_{ij}(\zeta ), 
\nonumber \\
\tilde{\chi}_{ij}(\zeta )=\chi_{ij}(\zeta )+\beta_{i,lk}D_{lknm}(\zeta )\beta_{j.nm}.
\label{pi8}
\end{eqnarray}
Eq.(\ref{pi7}) is the zero frequency limit \begin{math} \zeta \to i0^+ \end{math} of 
Eq.(\ref{pi8}). The dynamical dielectric response tensor  
\begin{math} \varepsilon_{ij}(\zeta ) \end{math} appears in the polarization part of the 
photon propagator\cite{Abrikosov:1963}. The Feynman diagrams contributing to the 
polarization part of the photon propagator in a piezoelectric system are shown in 
Fig. \ref{fig2}. These are equivalent to Eq.(\ref{pi8}) and explain why mechanical 
acoustic frequencies appear in the electrical response of piezoelectric 
materials. 

\section{Fracture and Stress \label{fs}}

\begin{figure}
\scalebox {0.5}{\includegraphics{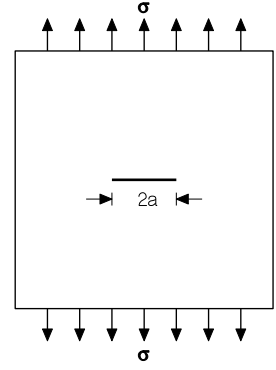}}
\caption{A micro-crack is formed in solid under stress $\sigma $. The 
width of the micro-crack is $2a$ and the length (into the paper) of the 
crack is $L\gg a$. The half width $a$ is the critical length size for forming 
the micro-crack as in Eq.(\ref{fs2}).} 
\label{fig3}
\end{figure}

Shown in FIG. \ref{fig3} is a crystal under stress \begin{math} \sigma \end{math} 
inducing a micro-crack of width \begin{math} 2a \end{math} and length 
\begin{math} L\gg a \end{math}. The energy 
\begin{math} U_b \end{math} required to create a micro-crack of half width 
\begin{math} b \end{math} and length \begin{math} L \end{math} is 
given by\cite{Landau:1970}
\begin{equation}
u(b)=\frac{U_b}{L}=4\gamma_s\ b -
\pi \left[\frac{(1-\nu^2)\sigma^2 }{\cal E}\right] b^2, 
\label{fs1}
\end{equation}   
wherein \begin{math} \gamma_s \end{math} is the surface tension of the 
micro-crack interface, \begin{math} {\cal E}  \end{math} is the material 
Young's modulus and \begin{math} \nu \end{math} is the Poisson ratio.

\subsection{Tensile Strength \label{ts}}

The maximum of the elastic micro-crack energy per unit length  
(\begin{math} \max_{b>0}\ u(b) \end{math}) represents the energy barrier 
to micro-crack creation. In detail,
\begin{eqnarray}
u=\max_{b>0}\ u(b)\ \ {\rm at}\ \  b=a,
\nonumber \\
a=\frac{2\gamma_s}{\pi }\left[\frac{\cal E}{(1-\nu^2)\sigma_F ^2 }\right],
\nonumber \\ 
u=2\gamma_s a = 
\frac{4\gamma_s ^2 }{\pi }\left[\frac{\cal E}{(1-\nu^2)\sigma_F ^2 }\right].
\label{fs2}
\end{eqnarray}   
The stress level \begin{math} \sigma_F \end{math} which nucleates a micro-crack 
is thereby the well known result\cite{LandL}
\begin{equation}
\sigma_F=\sqrt{\frac{2\gamma_s {\cal E}}{\pi (1-\nu ^2)a}}
\label{fd3}
\end{equation}
The tensile strength \begin{math} \sigma_F \end{math} of the material is then 
given by Eq.(\ref{intro4}) wherein the broken chemical bond strength  
\begin{equation}
\sigma_{\rm bond}=\frac{2{\cal E}}{\pi (1-\nu ^2)}
\label{fd4}
\end{equation} 
is determined by Young's modulus \begin{math} {\cal E}  \end{math} and 
the Poisson ratio \begin{math} \nu  \end{math}.

\subsection{Numerical Estimates \label{ne}}

Employing the values of material constants for fused quartz, we can estimate at least 
the powers of ten that would apply to piezoelectric rocks such as granite 
rocks. The values are 
\begin{eqnarray}
\gamma_s \sim 10^2 \ \frac{\rm erg}{\rm cm^2}\ , 
\nonumber \\ 
\sigma_{\rm bond}\sim 10^{12}\ \frac{\rm erg}{\rm cm^3}\ ,
\nonumber \\ 
\sigma_F \sim 10^9\ \frac{\rm erg}{\rm cm^3}\ ,
\nonumber \\ 
a\sim 10^{-4}\ {\rm cm}\ ,
\label{ne1}
\end{eqnarray}
in satisfactory agreement with the elastic theory as reviewed in Sec.\ref{ts}.

Some comments are in order: (i) For quartz, the value of 
\begin{math} a\sim 1\ {\rm micron}  \end{math}. (ii) For the brittle fracture 
of quartz, the macroscopic fracture surface experimentally exhibits micro-cracks with a length  
\begin{math} L\sim 20\ {\rm micron} \gg a \end{math}. (iii) As is usual in fractures 
\begin{math} \sigma_F \ll \sigma_{\rm bond} \end{math}, i.e. 
\begin{math} \sigma_F \sim 10^{-3} \sigma_{\rm bond} \end{math} for the problem at hand. 
(iv) The velocity of sound \begin{math} v_s  \end{math} compared with the velocity of light 
\begin{math} c \end{math} obeys \begin{math} (v_s/c)\sim 10^{-5} \end{math}. The ratio of 
phonon frequencies to photon frequencies in cavities of similar length scales thereby obey 
\begin{equation}
\left(\frac{\omega_{\rm phonon}}{\omega_{\rm photon}}\right) \sim 10^{-5}\ \ \ 
{\rm for\ similar\ sized \ cavities}.
\label{ne2}
\end{equation} 
The importance of the above Eq.(\ref{ne2}) is that the phonon modes enter into the dynamic 
dielectric response function \begin{math} \varepsilon (\omega +i0^+) \end{math} in virtue 
of Eq.(\ref{pi8}).

\section{Neutron Production \label{np}}

\begin{figure}
\scalebox {0.6}{\includegraphics{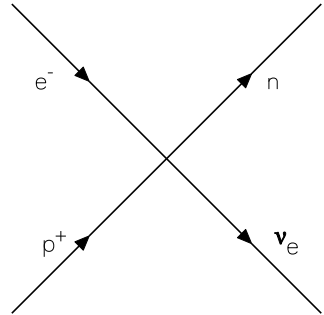}}
\caption{Neutron production takes place via the standard Fermi weak interaction 
as shown above. The electron energy is renormalized $mc^2\to W$ by condensed 
matter microwave radiation present as the stress approaches the fracture value $\sigma_F$. 
The coupling strength at the four Fermion vertex is $G_F$.} 
\label{fig4}
\end{figure}

The neutron production rate at the fracture stress \begin{math} \sigma_F \end{math} is here 
considered due to energetic electrons scattering off protons which are naturally present in 
(say) granite  as water or organic molecules. The Feynman diagram in the Fermi theory limit 
of the standard model is shown in FIG. \ref{fig4} described in Eq.(\ref{intro1}).

\subsection{Electron Renormalized Energy W \label{re}} 

To begin to analyze the production of neutrons via the reaction Eq.(\ref{intro1}), one must 
calculate the mean energy of electrons in condensed matter when accelerated by an electric field 
\begin{equation}
\frac{d{\bf p}}{dt}=e{\bf E}.
\label{np1}
\end{equation}
The electron energy is estimated to be 
\begin{equation}
W=\sqrt{m^2c^4+c^2\overline{|{\bf p}|^2}}.
\label{np2}
\end{equation}
If \begin{math} P_{\bf E}(\omega )d\omega \end{math} represents the the mean squared electric 
field strength in a bandwidth \begin{math} d\omega \end{math}, then Eq.(\ref{np1}), implies 
\begin{eqnarray}
E^2=\int_0^\infty P_{\bf E}(\omega ) d\omega,
\nonumber \\
\overline{|{\bf p}|^2}=e^2\int_0^\infty P_{\bf E}(\omega ) \frac{d\omega }{\omega^2 }.
\label{np3}
\end{eqnarray}
If \begin{math} \Omega  \end{math} denotes the dominant frequency in the spectrum of electric field 
fluctuations, then Eq.(\ref{np3}) is more simply written 
\begin{equation}
\overline{|{\bf p}|^2}=\frac{e^2E^2}{\Omega^2}
\label{np4}
\end{equation}
so that the ratio of the energy to the rest energy of the electron is 
\begin{equation}
\beta =\frac{W}{mc^2}=\sqrt{1+\left(\frac{eE}{mc\Omega}\right)^2}\ .
\label{np5}
\end{equation}
The value of \begin{math} \beta >1  \end{math} is critical for measuring whether or not there is 
sufficient radiation energy to allow for the reaction in Eq.(\ref{intro1}).  

\subsection{Further Numerical Estimates at Fracture \label{neaf}}

To estimate the electric field, one notes that the stress at fracture \begin{math} \sigma_F  \end{math} 
is in large part due to the  electric field strength 
\begin{equation}
\sigma_F \sim \frac{E^2}{4\pi}\ \ \ \Rightarrow 
\ \ \ E\sim 10^5\ {\rm Gauss},
\label{neaf1}
\end{equation}
in virtue of Eq.(\ref{ne1}). 
Since 
\begin{equation}
\frac{e}{mc}\approx 1.75882915 \times 10^7 \ \frac{1}{\rm Gauss\ second}\ ,
\label{neaf2}
\end{equation}
one finds 
\begin{equation}
\frac{eE}{mc}\sim \frac{10^{12}}{\rm second}\ . 
\label{meaf3}
\end{equation}
The frequency of a sound mode localized on a micro-crack of width 
\begin{math} 2a \end{math} for a reasonable sound velocity in rock is 
in the microwave range 
\begin{equation}
\Omega \sim \frac{10^9}{\rm second}
\label{meaf4}
\end{equation}
One should then observe electromagnetic microwave emission when the sound mode is turned into 
an electromagnetic mode via the piezoelectric effect. 

In virtue of Eqs.(\ref{np5}), (\ref{meaf3}) 
and (\ref{meaf4}) one finds \begin{math} \beta \sim 30 \end{math}. The threshold value of 
\begin{math} \beta  \end{math} for Eq.(\ref{intro1}) to be possible without radiation is  
\begin{math} \beta_0\approx 2.53  \end{math} so that the energy renormalized by radiation 
is above threshold by a wide margin, i.e. \begin{math} \beta \gg \beta_0 \end{math}.
The electron energies on the surface of a micro-crack in a stressed environment with an external 
stress \begin{math} \sigma_F \end{math} obey  
\begin{equation}
W \sim 15\ {\rm MeV}\ .
\label{meaf5}
\end{equation}    
The transition rate per unit time for Eq.(\ref{intro1}) by the usual standard has been 
computed\cite{Widom:2006,Srivastava:2010} as 
\begin{eqnarray}
\Gamma(e^- + p^+ \to n + \nu_e)\approx \left(\frac{G_F m^2}{\hbar c}\right)^2 \left(\frac{mc^2}{\hbar}\right)\beta^2,
\nonumber \\ 
\Gamma(e^- + p^+ \to n + \nu_e)\approx 7\times 10^{-3}\beta^2 \ {\rm Hz},
\nonumber \\ 
\Gamma(e^- + p^+ \to n + \nu_e) \sim  0.6   \ {\rm Hz} \ \ \ {\rm for}\ \ \ \beta \sim 30.
\label{meaf6}
\end{eqnarray}
The transition rate per unit time per unit area of micro-crack surfaces may be found from 
\begin{equation}
\varpi_2=n_2 \Gamma(e^- + p^+ \to n + \nu_e)
\label{meaf7}
\end{equation}
wherein \begin{math} n_2 \end{math} is the number of protons per unit micro-crack area in the first few layers of the quartz granite. Typical values are 
\begin{equation}
n_2\sim 2\times 10^{14}\ \frac{1}{\rm cm^2} \ \ \ \Rightarrow 
\ \ \ \varpi_2 \sim  10^{15} \frac{\rm Hz}{\rm cm^2}\ . 
\label{neaf8}
\end{equation}
If the fracture takes place with hydraulic fracture processes, then the neutron production rate will 
be about a factor of ten higher due to the higher water concentration on the micro-crack surface 
areas.

\section{Conclusions \label{conc}}

It is in the nature of piezoelectric matter that strong mechanical disturbances give rise to strong 
electromagnetic responses. This is true for piezoelectric rocks such as granite which contain large 
amounts of quartz. For large scale piezoelectric rock fracturing, as takes place in earthquakes, 
electromagnetic responses in many frequencies, from radio frequaecy to gamma ray frequancy, are to be 
expected. Some have attributed earthquake lights and/or lightning\cite{Ikeya:1996} to the phenomena 
discussed in this work.

We have employed the standard model of weak interactions along with the known theory 
of piezoelectric materials to explain the experimental evidence that  
fracturing  piezoelectric rocks produces neutrons. We have also explained why such 
fracturing processes produce microwave radiation. The elastic energy micro-crack 
production ultimately yields the macroscopic fracture whose acoustic vibrations are 
converted into electromagnetic oscillations. The electromagnetic microwaves 
accelerate the condensed matter electrons which then scatter from protons to produce 
neutrons and neutrinos. This work also
may have implications for a better understanding of radiative
processes associated with earthquakes\cite{Freund:2003,Takaki:1998}.


\begin{thebibliography}{11}

\bibitem{Karassey:1953}
V.V. Karassey, N.A. Krotova and B.W. Deryagin, {\it Dokl. Akad. Nauk SSSR} {\bf 88}, 777 (1953).

\bibitem{Klyuev:1986}  
V.A. Klyuev, A.G. Lipson, Yu.P. Toporov, B.V. Deryagin, V.J. Lushchikov, 
A.V. Streikov, E.P. Shabalin, {\it Sov. Tech. Phys. Lett.} {\bf 12}, 551 (1986).

\bibitem{Klyuev:1987}  
V. Klyuev et al., {\it Kolloidn. Zh.} {\bf 88}, 1001 (1987).

\bibitem{Preparata:1991} 
G. Preparata, {\it Il Nuovo Cimento} {\bf 104}, 1259 (1991)

\bibitem{Nakayama:1992} 
K. Nakayama, N. Suzuki and H. Hashimoto, 
{\it Journal of Physics} {\bf D 25}, 303 (1992)

\bibitem{Lawn:1993} 
B. Lawn, ``{\it Fracture of Brittle Solids}', Sec. 4.5, page 103,  
Cambridge University Press, Cambridge (1993)

\bibitem{Derjaguin:1989}
B.V. Derjaguin, A.G. Lipson, V.A. Kluev, D.M. Sakov and Yu.P. Toporov,
{\it Nature}, {\bf 341} , 492 (1989)

\bibitem{Lipson:1989} 
A. G. Lipson, D. M. Sakov, V. A. Klyuev and B. V. Deryagin, {\it  JETP. Lett.} {\bf 49}, 675 (1989). 

\bibitem{Derjaguin:1990} B.V. Derjaguin, V.A. Kluev, A.G. Lipson, and Yu.P Toporov,  
 {\it Physica} {\bf B 167}, 189 (1990).

\bibitem{Kaushik:1997}
T. Kaushik et al, {\it Phys. Lett.} {\bf A 232}, 384 (1997).

\bibitem{Shioe:2009} 
Y. Shioe et al., {\it Il Nuovo Cimento} {\bf 112} 1059 (1999).

\bibitem{Cardone:2009}
F. Cardone, A. Carpinteri, and G. Lacidogna, 
{\it Phys. Lett.} {\bf A 373}, 862 (2009).

\bibitem{Carpinteri:2009} 
A. Carpinteri and G. Lachidogna, {\it Strain} {\bf 45}, 332 (2009)
 
\bibitem{Fuji:2002} 
M. Fuji et al., {\it Jpn. J, Appl. Phys}, {\bf 41},  2115 (2002)

\bibitem{Landau:1984}
L.D. Landau and E.M. Lifshitz, {\it Electrodynamics of Continuous Media}, 
Sec. 17, Pergamon Press, Oxford (1984).

\bibitem{Fruend:1998}
L.B. Fruend, {\it Dynamic Fracture Mechanics}, Cambridge University Press, Cambridge (1998).

\bibitem{Landau:1970}
L.D. Landau and E.M. Lifshitz, {\it Theory of Elasticity}, 
Sec. 31, Pergamon Press, Oxford (1970). 

\bibitem{Griffith:1921}
A.A. Griffith, {\it Proc. Roy. Soc.} {\bf 221}, 161 (1921).

\bibitem{Abrikosov:1963}
A.A. Abrikosov, Gorkov and I.E. Dzyaloshinskii, 
{\it Qunatum Field Theory Methods in Statistical Physics}, Chapter 6,
Prentice Hall, Englewood Cliffs (1963).

\bibitem{LandL}
L.D. Landau and E.M. Lifshitz, {\it op. cit.} \cite{Landau:1970}, 
page 146, Eq.(31.10). 

\bibitem{Widom:2006}
A.Widom and L. Larsen {\it Eur. Phys. J.} {\bf C 46}, 107 (2006).

\bibitem{Srivastava:2010}
Y.N. Srivastava, A. Widom and L. Larsen, {\it Pramana } {\bf 75} 617 (2010). 

\bibitem{Ikeya:1996}
M. Ikeya and S. Takaki, {\it Jpn. J. Appl. Phys.} {\bf 35}, L355 (1996). 

\bibitem{Freund:2003}
F.T. Freund, ``{\it Rocks that Crackle and
Sparkle and Glow: Strange Pre-Earthquake Phenomena}'',
Journal of Scientific Exploration {\bf 17}, no. 1, p. 37-71 (2003).

\bibitem{Takaki:1998}
S. Takaki and M. Ikeya, ``{\it A Dark Discharge
Model of Earthquake Lightning}'', Japanese Journal of
Applied Physics {\bf 37} Issue 9A, 5016 (1998).



\end{thebibliography}
\end{document}